\documentclass[11pt,twoside]{article}
\usepackage{amsmath}
\usepackage{amssymb}
\usepackage{amscd}
\usepackage{mathrsfs}
\usepackage{graphicx}
\usepackage{epstopdf}
\usepackage[usenames]{color}
\usepackage{wrapfig}
\usepackage{boxedminipage}
\newtheorem{lemma}{Lemma}[section]
\newtheorem{theorem}{Theorem}[section]
\newcommand{\remark}{\smallbreak\noindent{\bf Remark.}}
\newcommand{\de}{\partial}
\newcommand{\dde}[2]{\frac{\partial #1}{\partial #2}}
\newcommand{\na}{\nabla\!}
\newcommand{\nasl}{{\rlap{\raise1pt\hbox{\,/}}\nabla}}
\newcommand{\codiv}{\nabla\!{\cdot}}
\newcommand{\dO}{\mathrm{d}}
\newcommand{\DO}{\mathrm{D}}
\newcommand{\jO}{\mathrm{j}}
\newcommand{\JO}{\mathrm{J}}
\newcommand{\LO}{\mathrm{L}}

\newcommand{\TO}{\mathrm{T}}
\newcommand{\TS}{\TO^{*}\!}
\newcommand{\VO}{\mathrm{V}}
\newcommand{\VS}{\VO^{*}\!}
\newcommand{\iO}{\mathrm{i}}
\newcommand{\End}{\operatorname{End}}
\newcommand{\Aut}{\operatorname{Aut}}
\newcommand{\Tr}{\operatorname{Tr}}
\newcommand{\into}{\hookrightarrow}
\newcommand{\onto}{\rightarrowtail}
\newcommand{\bang}[1]{{\langle#1\rangle}}
\newcommand{\fnb}[1]{[\![#1]\!]}
\newcommand{\pp}{{\mathsf{p}}}
\newcommand{\xx}{{\mathsf{x}}}
\newcommand{\yy}{{\mathsf{y}}}
\newcommand{\dx}{\dO\xx}
\newcommand{\dy}{\dO\yy}
\newcommand{\Cs}{{\rlap{\lower3pt\hbox{\textnormal{\LARGE \char'040}}}{\Gamma}}{}}
\newcommand{\C}{{\boldsymbol{C}}}
\newcommand{\E}{{\boldsymbol{E}}}
\newcommand{\Ee}{{\scriptscriptstyle{\boldsymbol{E}}}}
\newcommand{\F}{{\boldsymbol{F}}}
\newcommand{\Gb}{{\boldsymbol{G}}}
\newcommand{\Gg}{{\scriptscriptstyle{\boldsymbol{G}}}}
\newcommand{\M}{{\boldsymbol{M}}}
\newcommand{\Mm}{{\scriptscriptstyle{\boldsymbol{M}}}}
\newcommand{\W}{{\boldsymbol{W}}}
\newcommand{\Z}{{\boldsymbol{Z}}}
\newcommand{\Lie}{\mathfrak{L}}
\newcommand{\RR}{{\mathbb{R}}}
\newcommand{\Id}[1]{{1\!\!1}\!{}_{#1}{}}
\newcommand{\id}{{1\!\!1}}
\newcommand{\Ccal}{{\mathcal{C}}}
\newcommand{\Ecal}{{\mathcal{E}}}
\newcommand{\Jcal}{{\mathcal{J}}}
\newcommand{\Lcal}{{\mathcal{L}}}
\newcommand{\Pcal}{{\mathcal{P}}}
\newcommand{\Tcal}{{\mathcal{T}}}
\newcommand{\Ucal}{{\mathcal{U}}}
\newcommand{\ten}[1]{\operatorname*{\otimes}_{\!{\scriptscriptstyle #1}} }

\newcommand{\cart}[1]{\operatorname*{\times}_{\!{\scriptscriptstyle #1}} }
\newcommand{\dir}[1]{\operatorname*{\oplus}_{\!{\scriptscriptstyle #1}} }
\newcommand{\we}{{\,\wedge\,}}
\newcommand{\weu}[1]{{\wedge^{\!#1}}}
\newcommand{\pint}{{\scriptscriptstyle\mathord{\rfloor}}}
\newcommand{\comp}{\mathbin{\raisebox{1pt}{$\scriptstyle\circ$}}}
\newcommand{\tn}{{\,\otimes\,}}
\newcommand{\ul}{\underline}
\newcommand{\ost}[1]{\overset{{}_{{\,}_*}}{#1}}
\newcommand{\sst}{\scriptscriptstyle}
\newcommand{\up}{{\scriptscriptstyle\uparrow}}
\newcommand{\Rq}{{\Rightarrow\quad}}
\newcommand{\qR}{{\quad\Rightarrow}}
\newcommand{\spec}[1]{{}_{\scriptscriptstyle{\mathrm{#1}}}}
\newcommand{\rdg}{{\textstyle\sqrt{{\scriptstyle|}g{\scriptstyle|}}\,}}
\newcommand{\rrdg}{{\scriptstyle\sqrt{{\sst|}g{\sst|}}}}
\newcommand{\lfr}{\mathfrak{l}}
\newcommand{\Ii}[2]{{}^{#1}_{\phantom{#1}\!#2}}
\newcommand{\iI}[2]{{}_{#1}^{\phantom{#1}\!#2}}
\newcommand{\iIi}[3]{{}_{#1\phantom{#2}\!\!#3}^{\phantom{#1}\!#2}}
\newcommand{\IiI}[3]{{}^{#1\phantom{#2}\!\!#3}_{\phantom{#1}\!#2}}
\newcommand{\sco}[3]{{\mathsf{c}}\Ii{#1}{#2#3}}
\newcommand{\sH}{{\scriptscriptstyle H}}
\newcommand{\sI}{{\scriptscriptstyle I}}
\newcommand{\sJ}{{\scriptscriptstyle J}}
\newcommand{\oh}{\tfrac{1}{2}}
\newcommand{\ih}{\tfrac{\iO}{2}}
\newcommand{\oq}{\tfrac{1}{4}}
\newcommand{\iq}{\tfrac{\iO}{4}}
\renewcommand{\a}{\alpha}
\renewcommand{\b}{\beta}
\newcommand{\g}{\gamma}
\newcommand{\G}{\Gamma}
\renewcommand{\d}{\delta}
\renewcommand{\k}{\kappa}
\renewcommand{\l}{\lambda}
\newcommand{\m}{\mu}
\renewcommand{\r}{\rho}
\renewcommand{\t}{\tau}
\newcommand{\om}{{\omega}}
\newcommand{\sref}[1]{\S\ref{#1}}
\title{Overconnections and the energy-tensors \\
of gauge and gravitational fields}
\date{{\small version 2 - April 10, 2016} }
\author{Daniel Canarutto \\[6pt]
{\small\it Dipartimento di Matematica e Informatica ``U.~Dini'', }\\
{\small\it Via S. Marta 3, 50139 Firenze, Italia}\\
{\small email:~daniel.canarutto@unifi.it}\\
{\small http://www.dma.unifi.it/\char126 canarutto}}
\begin{document}
\maketitle \thispagestyle{empty}

\begin{abstract}\noindent
A geometric construction for obtaining a prolongation of a connection
to a connection of a bundle of connections is presented.
This determines a natural extension of the notion of canonical energy-tensor
which suits gauge and gravitational fields,
and shares the main properties of the energy-tensor of a matter field
in the jet space formulation of Lagrangian field theory,
in particular with regards to symmetries of the Poincar\'e-Cartan form.
Accordingly, the joint energy-tensor for interacting matter and gauge fields
turns out to be a natural geometric object,
whose definition needs no auxiliary structures.
Various topics related to energy-tensors,
symmetries and the Einstein equations
in a theory with interacting matter, gauge and gravitational fields
can be viewed under a clarifying light.
Finally, the symmetry determined by the ``Komar superpotential''
is expressed as a symmetry of the gravitational Poincar\'e-Cartan form.
\end{abstract}
\bigbreak

\bigbreak
\noindent
PACS 2010: 02.40.Hw, 04.20.Fy.

\bigbreak

\bigbreak
\noindent
DOI: 10.1016/j.geomphys.2016.03.027 

\vfill\newpage
\tableofcontents
\thispagestyle{empty}
\vfill\newpage
\setcounter{page}{1}

\section*{Introduction}

The so-called ``canonical energy-tensor'' relates ``basic'' vector fields,
that is vector fields on the source manifold,
to infinitesimal symmetries of the field theory under consideration
and currents associated with them.
While it is usually appreciated that the standard expression
\hbox{$\ell\,\d^a_b-\phi^i_{,b}\,\de^a_i\ell$}
has no geometric meaning on non-trivial bundles,
this issue is dealt with in various ways and with different formalisms,
sometimes with \emph{ad hoc} prescriptions.
On the whole we may say that issues related to energy-tensors,
including the stress-energy tensor in General Relativity,
continue to generate much interest and discussions in the literature~%
\cite{AndersonTorre94,BabakGrishchuk99,BambaShimizu15,ButcherLasenbyHobson08,
ChenNester00,FatibeneFerrarisFrancaviglia94,ForgerRoemer04,KijowskiGRG97,Leclerc06,
ObukhovPuetzfeld14,Pons09}.

It can be argued that the natural setting for a thorough clarification
of such matters is provided by the geometric formulation
of Lagrangian field theory on jet spaces and, in particular,
by generalizations of the Noether theorem expressed in terms
of symmetries of the Poincar\'e-Cartan form~\cite{FernandezGarciaRodrigo00,Gar74,
FerrarisFrancaviglia92,GS73,Kolar84,Krupka02,Krupka15,KrupkaKrupkovaSanders10,
Kuperschmidt79,MaMo83b,Tr67,VinogradovCSS84I,VinogradovCSS84II,Voicu2015}.
In that context, a previous paper by M.~Modugno and myself~\cite{CM85}
offered a natural extension of the notion of canonical energy-tensor
to the non-trivial bundle case,
on the basis of an earlier suggestion by Hermann~\cite{Hermann75}.

The basic notions related to the Poincar\'e-Cartan form
are reviewed here in \sref{ss:Poincare-Cartan form and currents};
the above said construction of the energy-tensor,
which uses a fixed background connection of the involved bundle, is reviewed in
\sref{ss:Canonical energy-tensor in non-trivial bundles}.
In \sref{ss:Basic examples} we discuss two basic examples and, in those contexts,
the canonical energy-tensor is compared to the stress-energy tensor
appearing in the right-hand side of the Einstein equations
when the considered field is coupled to gravity.

The extension of the above said approach to the case when the considered field
is a connection poses additional problems.
First we should note that a general connection of a bundle \hbox{$\E\onto\M$}
cannot be characterized as a section of a finite-dimensional bundle
over the base manifold $\M$.
However a smooth algebraic structure of the fibers
does select a special family of connections, which can be regarded
as the sections of a natural finite-dimensional bundle \hbox{$\C\onto\M$}.
This idea was first introduced by Garcia~\cite{Gar72} in the context
of principal bundles, and then generalized and used by Modugno
and others~\cite{DodsonModugno86,Modugno87b,
CabrasCanarutto91II,CanaruttoDodson85,Janyska07},
sometimes with the designation ``systems of connections''.
One interesting aspect of that formalism is that the bundle \hbox{$\C\onto\M$}
inherits a fibered algebraic structure,
which in turn determines a new system of connections of it,
called ``overconnections'' (or ``connections over connections'').

The needed notions related to systems of connections and overconnections
are introduced in~\sref{ss:Prolongations of general connections},
\sref{ss:Systems of linear connections and overconnections}
and~\sref{ss:Systems of gauge fields}.
In particular we consider linear connections of vector bundles.
In this context we explicitely describe the bundle \hbox{$\C\onto\M$}
and its sub-bundles determined by possible further fiber structures,
as well as the induced systems of overconnections.
Moreover we show how, via natural geometric constructions,
a linear connection can be lifted to an overconnection
with the aid of a linear connection of the base manifold.

This lift of connections to overconnections,
presented here for the first time as far as I know,
naturally yields the wanted definitions of energy-tensors for gauge fields
(\sref{ss:Energy-tensor of a gauge field}).
Though the proposed construction essentially arises by analogy
with the canonical energy-tensor of a matter field
described in~\sref{ss:Canonical energy-tensor in non-trivial bundles},
it is fully justified \emph{a posteriori} by its properties.
Indeed we find that the same relations among basic vector fields
and symmetries of the Poincar\'e-Cartan form still hold.
Furthermore the total energy-tensor
for a theory of interacting matter and gauge fields
on a gravitational background arises now very naturally,
and is independent of any auxiliary structures.
This total object (not the single pieces)
turns out to be divergence-free on-shell, as one would expect.

The further extension of these ideas to a theory
of coupled matter, gauge and gravitational fields is still quite natural
(\sref{ss:Energy-tensor of the gravitational field}).
Indeed the definition of energy-tensor for the gravitational field,
in the so-called ``metric-affine'' approach,
essentially follows from the same procedure
and fulfills the expected properties in relation to
the Poincar\'e-Cartan form.
In this enhanced geometric context we can view under a different light
issues and results discussed with special
emphasis by Padmanabhan~\cite{Padmanabhan1312.3253},
in particular about the relation among basic vector fields, energy-tensors,
symmetries and the Einstein equations.

In \sref{ss:The off-shell symmetry associated with any vector field}
we examine the off-shell symmetry of the gravitational field determined
by any vector field on the spacetime manifold~\cite{Komar59,Padmanabhan1312.3253}.
This turns out to be indeed related to a symmetry of the Poincar\'e-Cartan form
of the gravitational field,
obtained via a certain natural lift which is different,
and somewhat more complicated, than the horizontal lift
used in the definition of the energy-tensor.

\smallbreak\noindent
{\bf Acknowledgements.}~I wish to thank
Ettore Minguzzi, Marco Modugno and Raffaele Vitolo for discussions and remarks.

\section{Connections and overconnections}
\label{s:Connections and overconnections}

\subsection{Prolongations of general connections}
\label{ss:Prolongations of general connections}

According to a certain line of thought,
the basic ideas related to the notion of a connection are best
expressed in a general context in which no fiber structure
is assumed in the bundle under consideration.
A possible algebraic fiber structure enters the picture at a later step,
selecting a special family of connections.
Indeed, a connection of an arbitrary finite-dimensional smooth fibered manifold 
\hbox{$\E\onto\M$} is defined to be a (smooth) section \hbox{$\k:\E\to\JO\E$}
(we denote by $\TO$, $\VO$ and \hbox{$\JO\equiv\JO_1$} the tangent, vertical
and first-jet prolongation functors).
We recall that \hbox{$\pp_\Ee:\JO\E\onto\E$} is naturally an affine bundle, as
$$\mathrm{d\!l}:\JO\E\into\TS\M\ten{\E}\TO\E$$
is the sub-bundle over $\E$ which projects over the identity section
\hbox{$\M\to\TS\M\tn\TO\M$}.

The tangent prolongation \hbox{$\TO\phi:\TO\M\to\TO\E$}
of a local section \hbox{$\phi:\M\to\E$} can be regarded as a section
\hbox{$\M\to\TS\M\ten{\E}\TO\E$} and,
since it projects over the identity of $\TO\M$,
also as a section \hbox{$\jO\phi:\M\to\JO\E$}
(the \emph{first-jet prolongation} of $\phi$).
If $\bigl(\xx^a,\yy^i\bigr)$ is a local fibered coordinate chart of $\E$
then we denote by $\bigl(\xx^a,\yy^i,\yy^i_a\bigr)$
the induced fibered chart of $\JO\E$, namely we have
$$\yy^i_a\comp\jO\phi=\phi^i_{,a}\equiv\de_a\phi^i~.$$
The components of a connection $\k$ are the functions
$$\k^i_a\equiv\yy^i_a\comp\k:\E\to\RR~.$$
Since $\k$ can be regarded as a tangent-valued one-form on $\E$,
its \emph{curvature tensor} can be introduced,
in terms of the Fr\"olicher-Nijenhuis bracket,
as the vertical-valued two-form
$$\r:= -\fnb{\k,\k}=\r\iI{ab}i\,\dx^a\we\dx^b\tn\de\yy_i:
\E\to\weu2\TS\M\ten{\E}\VO\E$$
where
\hbox{$\r\iI{ab}i=\k^i_{a,b}-\k^i_{b,a}+\de_j\k^i_a\,\k^j_b-\de_j\k^i_b\,\k^j_a$}\,.

The \emph{vertical projection} associated with $\k$ is the vertical-valued 1-form
$$\om:=\Id{\TO\E}-\k\;:\;\E\to\TS\E\ten{\E}\VO\E~,$$
which yields the \emph{covariant derivative} of a section \hbox{$\phi:\M\to\E$} as
$$\nabla\phi:=\jO\phi\pint\om\;:\;\M\to\TS\M\ten{\E}\VO\E~.$$

We are now interested in seeing under which conditions
a connection $\k$ of \hbox{$\E\onto\M$} can be lifted
to a connection of \hbox{$\JO\E\onto\M$}.
We start by noting that
the first jet prolongation of \hbox{$\k:\E\to\JO\E$} is a morphism
$$\JO\k:\JO\E\to\JO\JO\E~.$$
This is not a connection of \hbox{$\JO\E\onto\M$},
because it is not a section of \hbox{$\pp_{\sst\JO\E}:\JO\JO\E\to\JO\E$}
but rather a section of \hbox{$\JO\pp_{\sst\E}:\JO\JO\E\to\JO\E$}.
We can express this argument in terms of local coordinates as follows.
We denote the induced coordinate chart of $\JO\JO\E$ by
$$\bigl(\xx^a,\yy^i,\ul\yy{}^i_a\,;\yy^i_a,\yy^i_{ab}\bigr)~,
\qquad \yy^i_{ab}\equiv(\yy^i_a)^{\phantom{i}}_b\neq\yy^i_{ba}~,$$
namely
$$\yy^i_a\comp\pp_{\sst\JO\E}=\ul\yy{}^i_a~,\qquad
\yy^i_a\comp\JO\pp_{\sst\E}=\yy^i_a~,$$
and obtain
$$\bigl(\xx^a,\yy^i,\ul\yy{}^i_a\,;\yy^i_a,\yy^i_{ab}\bigr)\comp\JO\k=
\bigl(\xx^a,\yy^i,\k^i_a\,;\yy^i_a,\k^i_{a,b}\,{+}\,\yy^j_b\,\de_j\k^i_a\bigr)~.$$
Thus we could turn $\JO\k$ into a connection of \hbox{$\JO\E\onto\M$}
if we availed of an involution of $\JO\JO\E$ exchanging
$\ul\yy{}^i_a$ and $\yy^i_a$\,.
However this generalization of the involution of the double tangent space of a manifold
requires some added structure, namely a symmetric linear connection $\G$
of \hbox{$\TO\M\onto\M$}.
Indeed it has been proved by Modugno~\cite{Modugno87a}
that $\G$ determines a distinguished involution
\hbox{$\mathrm{s}_{\sst\G}:\JO\JO\E\to\JO\JO\E$},
with the coordinate expression
(note the exchange \hbox{$\yy^i_{ab}\to\yy^i_{ba}$})
$$\bigl(\xx^a,\yy^i,\ul\yy{}^i_a\,;\yy^i_a,\yy^i_{ab}\bigr)\comp\mathrm{s}_{\sst\G}=
\bigl(\xx^a,\yy^i,\yy^i_a\,;\ul\yy{}^i_a,
\yy^i_{ba}+\G\iIi bca\,(\ul\yy{}^i_c-\yy^i_c)\bigr)~.$$

Using the above cited result we now easily prove:
\begin{theorem}\label{theorem:connection_prolongation}
The composition
$$\k'\equiv\mathrm{s}_{\sst\G}\comp\JO\k:\JO\E\to\JO\JO\E$$
is a connection of \hbox{$\JO\E\onto\M$} which is projectable over $\k$\,.
Its coordinate expression turns out to be
$$\bigl(\xx^a,\yy^i,\ul\yy{}^i_a\,;\yy^i_a,\yy^i_{ab}\bigr)\comp\k'=
\bigl(\xx^a,\yy^i,\yy^i_a\,;\k^i_a,
\k^i_{b,a}\,{+}\,\yy^j_a\,\de_j\k^i_b
\,{+}\,\G\iIi bca\,(\k^i_c\,{-}\,\yy^i_c) \bigr)~.$$
\end{theorem}

A somewhat more manageable form for
the coordinate expression of $\k'$ is
$$(\k'_a)^i=\k^i_a~,\qquad
(\k'_a)^i_b=\k^i_{a,b}\,{+}\,\yy^j_b\,\de_j\k^i_a
\,{+}\,\G\iIi acb\,(\k^i_c\,{-}\,\yy^i_c)~.$$
Moreover we remark that the projectability property of $\k'$ can be expressed
by the commutative diagram
$$\begin{CD}
\JO\E @>{\k'}>> \JO\JO\E \\
@V{\pp_\Ee}VV         @VV{\JO\pp_\Ee}V \\
\E@>>{\k}> \JO\E
\end{CD}$$

\subsection{Systems of linear connections and overconnections}
\label{ss:Systems of linear connections and overconnections}

In works by Modugno and others~\cite{DodsonModugno86,Modugno87b}
the term ``overconnection'' denotes a connection
of a finite-dimensional ``bundle of connections'',
arising in the context of ``system of connections''.
A generic connection \hbox{$\E\to\JO\E$}
cannot be characterized as a section of a finite-dimensional bundle over $\M$.
In many relevant cases, however,
one avails of a fibered algebraic structure of \hbox{$\E\onto\M$},
selecting a special class of connections that can be seen as sections
of some finite-dimensional bundle \hbox{$\C\onto\M$}.
More precisely, there is an ``evaluation morphism''
$$\chi:\E\cart{\M}\C\to\JO\E~,$$
such that every section \hbox{$c:\M\to\C$}
determines a special connection via the composition
$$\E~\xrightarrow{\displaystyle~(\mathrm{id}\,,\,c\comp\pp_{\Mm})~}~\E\cart{\M}\C
~\xrightarrow{\displaystyle~~~\chi~~}~\JO\E~,$$
and, conversely, every special connection is obtained in this way.
The map $\chi$ determines various structures on $\C$.
The ensuing theory of system of connections
(and, more generally, of systems of sections of 2-fibered bundles)
has been studied in various contexts.

In the literature, the notion of bundle of connections has been
mainly exploited in relation to principal connections in gauge field theories
(e.g.\ see Jany\v{s}ka~\cite{Janyska07} and the references therein).
The bundle of linear connections of a vector bundle provides
an alternative, viable point of view, which, as far as I know,
has not yet been thoroughly explored.

Let \hbox{$\E\onto\M$} be a vector bundle.
Then \hbox{$\JO\E\onto\M$} turns out to be a vector bundle too,\footnote{
While \hbox{$\JO\E\onto\E$} is still an affine bundle.} 
as the jet prolongation functor $\JO$ naturally lifts
the algebraic fiber structure
(the zero section and the vector space operations).
Accordingly, we say that a connection \hbox{$\k:\E\to\JO\E$} is \emph{linear}
if it is a linear morphism over $\M$.
In that case we can regard it as a section \hbox{$\M\to\JO\E\ten{\M}\E^*$}
projecting over the identity of $\E$.
In other words, we can regard any linear connection of $\E$
as a section \hbox{$\M\to\C$} where
$$\C\subset\JO\E\ten{\M}\E^*,$$
called the \emph{bundle of linear connections of \hbox{$\E\onto\M$}},
is the affine subbundle over $\M$ which projects over the identity section
\hbox{$\Id{\E}:\M\to\E\ten{\M}\E^*$}.
The associated vector bundle is\footnote{
In fact \hbox{$\VO\E\cong\E\cart{\M}\E$}
because \hbox{$\E\onto\M$} is a vector bundle.} 
$$\DO\C=\DO\JO\E\ten{\M}\E^*=\TS\M\ten{\M}\E\ten{\M}\E^*\equiv
\TS\M\ten{\M}\End\E~.$$

We now assume that the chosen fiber coordinates $\bigl(\yy^i\bigr)$ are linear.
Then we obtain induced coordinates $\smash{\bigl(\xx^a,\yy\Ii ij,\yy^i_{aj}\bigr)}$
on \hbox{$\JO\E\ten{\M}\E^*$},
so that $\C$ is the submanifold
locally characterized by the constraint \hbox{$\smash{\yy\Ii ij=\d^i_j}$}\,.
If \hbox{$c:\M\to\C$} is a section
then the induced linear connection\footnote{
In the present context it's usually safe to dispense with the explicit
use of the evaluation morphism $\chi$\,,
so as a rule we identify \hbox{$c:\M\to\C$} with
\hbox{$\k=\chi\comp(\mathrm{id}\,,\,c\comp\pp_{\Mm})$\,.} } 
$\k$ is locally characterized by the components
\hbox{$\smash{\k^i_a=\k\iIi aij\,\yy^j}$} with
\hbox{$\smash{\k\iIi aij\equiv\yy^i_{aj}\comp c}$}\,.

Similar arguments hold for an affine bundle \hbox{$\F\onto\M$},
namely we define the \emph{affine connections} of it\footnote{
The term ``affine connection'' here is not to be intended in the same sense
as in many physics texts, where it essentially relates to the fiber structure
of \hbox{$\JO\E\onto\E$}.} 
as the affine morphisms \hbox{$\F\to\JO\F$} which project over
the identity of $\F$.
Now, since the above introduced bundle \hbox{$\C\onto\M$} is affine,
it has in turn a distinguished system of connections,
called \emph{the natural system of overconnections of $\E$}.
Using theorem~\ref{theorem:connection_prolongation}
we can now show that a distinguished overconnection naturally arises
from objects that are available in a standard gauge field theory.
First, by a coordinate computation one easily proves:
\begin{lemma}\label{lemma:induced_linear_overconnection1}
Let $\k$ be a linear connection of \hbox{$\E\onto\M$}
and $\G$ a linear connection of \hbox{$\TO\M\onto\M$}.
Then \hbox{$\k'\equiv\mathrm{s}_{\sst\G}\comp\JO\k:\JO\E\to\JO\JO\E$}
is a linear connection of \hbox{$\JO\E\onto\M$}, whose components are
$$\begin{cases}
(\k'_a)^i=\k\iIi aij\,\yy^j~,\\[6pt]
(\k'_a)^i_b=\de_b\k\iIi aij\,\yy^j+\yy^j_b\,\k\iIi aij
+\G\iIi acb\,(\k\iIi cij\,\yy^j-\yy^i_c)~.
\end{cases}$$
\end{lemma}
Moreover standard arguments about induced connections
of tensor product bundles yield:
\begin{lemma}\label{lemma:induced_linear_overconnection2}
Let $\k$ and $\G$ be as in~\ref{lemma:induced_linear_overconnection1},
and let $\ost\k$ be the ``dual'' linear connection of \hbox{$\E^*\onto\M$}.
Then \hbox{$\k^\up\equiv\k'\tn\ost\k$} is a linear connection
of \hbox{$\JO\E\ten{\M}\E^*\onto\M$}, with coefficients
$$\begin{cases}
(\k^\up_a)\Ii ij=-\k\iIi ahj\,\yy\Ii ih+\k\iIi aih\,\yy\Ii hj~,
\\[6pt]
(\k^\up_a)^i_{bj}=\de_b\k\iIi aih\,\yy\Ii hj
-\k\iIi ahj\,\yy^i_{bh}+\yy^h_{bj}\,\k\iIi aih
+\G\iIi acb\,(\k\iIi cih\,\yy\Ii hj-\yy^i_{cj})~.
\end{cases}$$
\end{lemma}

Now from the observation that
$$\VO\C\subset\VO\bigl(\JO\E\ten{\M}\E^*\bigr)=
\bigl(\JO\E\ten{\M}\E^*\bigr)\cart{\M}\bigl(\JO\E\ten{\M}\E^*\bigr)$$
is the subbundle which projects over $\id\times0$ we find:
\begin{theorem}\label{theorem:induced_linear_overconnection}
The connection $\k^\up$ of~lemma~\ref{lemma:induced_linear_overconnection2}
is reducible to an affine connection of the subbundle \hbox{$\C\onto\M$},
whose coefficients turn out to be
$$(\k^\up_a)^i_{bj}=\de_b\k\iIi aij-\k\iIi ahj\,\yy^i_{bh}
+\yy^h_{bj}\,\k\iIi aih
+\G\iIi acb\,(\k\iIi cij-\yy^i_{cj})~.$$
\end{theorem}

Summarizing, $\k$ and $\G$ together determine an overconnection,
that is an affine connection of \hbox{$\C\onto\M$},
which we'll still denote as $\k^\up$.
In particular we may express the covariant derivative
of $\k$ itself with respect to $\k^\up$, getting
\begin{align*}
\na_a\k\iIi bij&=\de_a\k\iIi bij-\de_b\k\iIi aij
+\k\iIi bih\,\k\iIi ahj
-\k\iIi aih\,\k\iIi bhj=
\\[6pt]
&=-\r\iIi{ab}ij~.
\end{align*}
Note that the contribution of the spacetime connection $\G$ disappears
in the above expression.

\subsection{Systems of gauge fields}
\label{ss:Systems of gauge fields}

In most situations of interest the vector bundle \hbox{$\E\onto\M$}
is endowed with a richer fiber structure.
The linear connections which preserve that structure,
let's call them \emph{gauge fields}, constitute a \emph{subsystem};
namely, they can be characterized as sections of an affine sub-bundle
of the bundle \hbox{$\C\onto\M$}
introduced in \sref{ss:Systems of linear connections and overconnections}.

The bundle of all linear endomorphisms of $\E$
is \hbox{$\End\E\equiv\E\ten{\M}\E^*\onto\M$}.
Its fibers are endowed with a natural Lie algebra structure
given by the ordinary commutator.
We denote by $\Aut\E$ its sub-bundle of all invertible endomorphisms;
this is a \emph{group bundle}, and $\End\E$ is its ``Lie-algebra bundle''.
Let now \hbox{$\Gb\subset\Aut\E$} be the sub-bundle of all automorphims
which preserve the assigned fiber structure of $\E$.
Its Lie-algebra bundle is a sub-bundle \hbox{$\Lie\subset\End\E$},
and the bundle of gauge fields is easily recognized as the affine sub-bundle
\hbox{$\C\!_\Gg\subset\C$} over $\M$ whose associated vector bundle is
$$\DO\C\!_\Gg=\TS\M\ten{\M}\Lie\subset\TS\M\ten{\M}\End\E~.$$
Moreover, the curvature tensor of any \hbox{$\k:\M\to\C\!_\Gg$}
can be regarded as a section
$$\r:\M\to\weu2\TS\M\ten{\M}\Lie~.$$

If one chooses a special frame of $\E$,
coordinate calculations are essentially the same as in the familiar
principal bundle formalism.
Let's consider the common case in which the fiber is complex
with fiber dimension $n$\,,
and is endowed with a Hermitian metric:
then $\Lie$ is the bundle of all \emph{anti-Hermitian} endomorphisms
(\hbox{$A^\dag=-A$}).
As a real vector bundle of fiber dimension $2n^2$,
$\End\E$ is endowed with the distinguished symmetric bilinear form
\hbox{$(A,B)\mapsto\Re\Tr(A\comp B)$}\,,
whose signature turns out to be $(n^2,n^2)$.

Now the assigned Hermitian structure of $\E$ 
also yields the Hermitian structure of $\End\E$ given by
\hbox{$(A,B)\mapsto\Tr(A^\dag\comp B)$},
and the real splitting \hbox{$\End\E=\Lie\dir{\M}\iO\,\Lie$}
(any endomorphism can be uniquely written
as the sum of anti-Hermitian and Hermitian terms).
The restrictions of this Hermitian product to $\Lie$ and $\iO\,\Lie$
turn out to be real positive (Euclidean),
while the restrictions of the real scalar product have opposite signatures.

If $\bigl(\yy_i\bigr)$ is an orthonormal frame of $\E$
then the matrix of a section \hbox{$\M\to\Lie$} is anti-Hermitian.
In particular, one can always find
an orthonormal frame $\bigl(\lfr_\sI\bigr)$ of $\Lie$
related to $\bigl(\yy_i\bigr)$ by the relations
\hbox{$\lfr_\sI=\lfr\iIi\sI ij \,\yy_i\tn\yy^j$},
where the matrices $\bigl(\lfr\iIi\sI ij\bigr)$ are \emph{constant}.
We then obtain the constant coefficients (\emph{structure constants})
$$\sco\sI\sJ\sH\equiv\bang{\lfr^\sI,[\lfr_\sJ\,,\,\lfr_\sH]}~,$$
where $\bigl(\lfr^\sI\bigr)$ is the dual frame.
Indices can be lowered an raised via the above said positive metric of $\Lie$.

Accordingly, the coordinate expressions of a gauge field
and that of its curvature can be written as
$$\k=\dx^a\tn(\de\xx_a+\k_a^\sI\,\lfr_\sI)~,\qquad
\r=\r\iI{ab}\sI\,\dx^a\we\dx^b\tn\lfr_\sI~,$$
with \hbox{$\r\iI{ab}\sI=
\k^\sI_{a,b}-\k^\sI_{b,a}+\sco\sI\sJ\sH\,\k_a^\sJ\,\k_b^\sH$}\,.
(In order to compare with the usual physics literature write
\hbox{$\k_a^\sI\equiv\iO\,q\,A_a^\sI$}
and \hbox{$\r_{ab}^\sI\equiv\iO\,q\,F_{ab}^\sI$} with \hbox{$q\in\RR$}\,.)

\smallbreak
The \emph{dual connection} $\ost\k$ and its curvature $\ost\r$
can be viewed as valued in the dual Lie algebra bundle
\hbox{$\Lie^*\subset\End\E^*\equiv\E^*\tn\E$},
and are related to $\k$ and $\r$ by \emph{minus transposition}:
in terms of fiber indices of $\E$ we have
$$\ost\k\iI{aj}i=-\k\iIi aij~,\qquad \ost\r\iI{abj}i=-\k\iIi{ab}ij~.$$
In terms of Lie algebra indices we'll write their coordinate expressions
as $\ost\k_{a\sI}$ and $\ost\r_{ab\sI}$\,;
the operations \hbox{$\k_a^\sI\mapsto\ost\k_{a\sI}$}
and \hbox{$\r_{ab}^\sI\mapsto\ost\r_{ab\sI}$} can also be regarded as
\emph{index lowering} with respect to the above said Hermitian structure of $\End\E$.

Recalling the results of~\sref{ss:Systems of linear connections and overconnections}
we now make the main point of this section,
which is easily proved by inspecting the general expression
of the overconnection $\k^\up$ determined by $\k$
with the aid of a suitable connection on the base manifold.
\begin{theorem}\label{theorem:gauge_overconnection}
If we avail of a linear torsion-free connection $\G$ of \hbox{$\TO\M\onto\M$},
then any gauge field $\k$ determines a connection $\k^\up$ of \hbox{$\C\onto\M$}
which turns out to be reducible to a connection of \hbox{$\C\!_\Gg\onto\M$},
with the coordinate expression
$$(\k^\up_a)^\sI_b=\k^\sI_{a,b}+\sco\sI\sJ\sH\,\k_a^\sJ\,\yy_b^\sH
+\G\iIi acb\,(\k^\sI_c-\yy^\sI_c)~.$$
\end{theorem}

As in the ``unconstrained'' case,
the covariant derivative of $\k$ with respect to $\k^\up$ is independent of $\G$
and has the  coordinate expression
$$\na_a\k_b^\sI=-\r\iI{ab}\sI~.$$
The above result is related to the ``Utiyama theorem''
and its generalizations~\cite{Janyska07},
which we'll comment about further in~\sref{ss:Energy-tensor of a gauge field}.

\section{Energy-tensor in Lagrangian field theories}
\label{s:Energy-tensor in Lagrangian field theories}

\subsection{Poincar\'e-Cartan form and currents}
\label{ss:Poincare-Cartan form and currents}

We recall a few basic notions in Lagrangian field theory.
A \emph{1-st order Lagrangian density} on a fibered manifold \hbox{$\E\onto\M$}
is defined to be a morphisms \hbox{$\Lcal:\JO\E\to\weu{m}\TS\M$} over $\M$,
where \hbox{$m\equiv\dim\M$}.
We write its coordinate expression as \hbox{$\Lcal=\ell\,\dO^m\xx$}\,,
with \hbox{$\dO^m\xx\equiv\dx^1\we{\cdot}{\cdot}{\cdot}\we\dx^m$}.
The associated \emph{Euer-Lagrange operator}
$$\Ecal:\JO_2\E\to\weu{m}\TS\M\ten{\E}\VS\E$$
has the coordinate expression\footnote{
If \hbox{$f:\JO\E\to\RR$} then the functions
\hbox{$\dO_af=\de_af+\yy^i_a\,\de_if+\yy^i_{ab}\,\de^b_if:\JO_2\E\to\RR$}
are the components of its \emph{horizontal differential},
see e.g.\ Saunders~\cite{Sa89}.} 
\hbox{$\Ecal_i=\de_i\ell-\dO_a\de^a_i\ell$}\,;
critical sections \hbox{$\phi:\M\to\E$} are characterized by the condition
\hbox{$\Ecal\comp\jO_2\phi=0$}\,.

As \hbox{$\JO\E\onto\E$} is an affine bundle and
its associated vector bundle is \hbox{$\DO\JO\E=\TS\M\ten{\E}\VO\E$},
the \emph{fiber derivative} of $\Lcal$ has a well-defined meaning as a morphism
$$\DO\Lcal:\JO\E\to(\DO\JO\E)^*\ten{E}\weu{m}\TS\M=
\TO\M\ten{\E}\VS\E\ten{E}\weu{m}\TS\M~.$$
We can transform this object via some natural operations:
we perform an obvious contraction and anti-symmetrization,
and use the transpose of the \emph{contact 1-form}\footnote{
This is the natural morphism over $\E$ with coordinate expression
\hbox{$\vartheta^i\equiv\dy^i\comp\vartheta=\dy^i-\yy^i_a\,\dx^a$}
~\cite{MaMo83b}.} 
\hbox{$\vartheta:\TO\JO\E\to\VO\E$}.
We end up with an $m$-form
$$\Pcal=\Pcal^a_i\,\vartheta^i\we\dx_a\equiv
\de^a_i\ell\,(\dy^i-\yy^i_b\,\dx^b)\we\dx_a\;:\;\JO\E\to\weu{m}\TO^*\JO\E~,$$
which by analogy with mechanics is sometimes called ``momentum''.
Since we have natural inclusions
\hbox{$\TS\M\subset\TS\E\subset\TO^*\JO\E$},
the \emph{Poincar\'e-Cartan form} \hbox{$\Ccal\equiv\Lcal+\Pcal$}
is a well-defined $m$-form on $\JO\E$, too.

In the present context we deal with symmetries of the Poincar\'e-Cartan form
rather than symmetries of the action functional
(the latter is the most common way in which these matters are formulated
in the physics literature).
A \emph{symmetry of $\Ccal$} is a vector field \hbox{$Z:\JO\E\to\TO\JO\E$}
such that the Lie derivative \hbox{$\LO_Z\Ccal$} vanishes along
all critical sections,\footnote{
If $\a$ is any $q$-form on $\JO\E$ then for any section \hbox{$\phi:\M\to\E$}
the pull-back \hbox{$\jO\phi^*\a$} is a $q$-form on $\M$.
In particular one has \hbox{$\dO\jO\phi^*\a=\jO\phi^*\dO\a$}\,.
} 
that is \hbox{$\jO\phi^*\LO_Z\Ccal=0$}\,.

The above definition can be refined via the following observations.
Let \hbox{$Y:\JO\E\to\TO\E$} be a morphism over $\E$.
By noting that for any section $\phi$ one has \hbox{$\jO\phi^*\vartheta^i=0$},
it is not difficult to check that
the forms \hbox{$\jO\phi^* i_Z\Ccal$}\,,
\hbox{$\jO\phi^* i_Z\dO\Ccal$}
and \hbox{$\jO\phi^*\LO_Z\Ccal$}
are independent of the choice of a vector field \hbox{$Z:\JO\to\TO\JO\E$}
such that \hbox{$\pp_{\!\Ee}\comp Z=Y$}.
Accordingly we say that $Y$ is a symmetry of $\Ccal$ if
\hbox{$\jO\phi^*\LO_Z\Ccal=0$}
for any such extension $Z$ and for any critical section $\phi$\,.

A \emph{conserved current} is defined to be an $m\,{-}\,1$-form
\hbox{$\Jcal:\JO\E\to\weu{m-1}\TO^*\JO\E$} such that the $m$-form
$\jO\phi^*\dO\Jcal$ on $\M$ vanishes for any critical section $\phi$\,.
It can be proved that the condition that $\phi$ be critical
can be equivalently expressed as \hbox{$(\jO\phi)^*(i_Z\dO\Ccal)=0$}
for any vector field $Z$\,;
then one immediately proves the following generalized version
of the Noether theorem:

\begin{theorem}~\\
$\bullet$~If \hbox{$Y:\JO\E\to\TO\E$} is a symmetry of $\Ccal$
then \hbox{$i_Y\Ccal:\JO\E\to\weu{m{-}1}\TO^*\JO\E$}
is a conserved current.\smallbreak\noindent
$\bullet$~If there exists an $m\,{-}\,1$-form
\hbox{$\varphi:\JO\E\to\weu{m-1}\TO^*\JO\E$} such that
\hbox{$\jO\phi^*\LO_Z\Ccal=\jO\phi^*\dO\varphi$}\,,
then, more generally, \hbox{$i_Y\Ccal\,{-}\,\varphi$} is a conserved current.
This condition is locally equivalent to
\hbox{$\jO\phi^*\LO_Z\dO\Ccal=0$}\,.
\end{theorem}

For any \hbox{$Y:\JO\E\to\TO\E$} a simple calculation yields
\begin{align*}
\jO\phi^*(i_Y\Ccal)&=
\jO\phi^*\bigl(i_Y\Lcal+i_{Y-Y\pint\mathrm{d\!l}}\Pcal\bigr)=
\\[6pt]
&=\bigl(\ell\,Y^a+\Pcal^a_i\,(Y^i-Y^b\,\phi^i_{,b})\bigr)\,\dx_a~.
\end{align*}
From this we see that there is a possible situation in which
one easily finds a symmetry $Y$ yielding a given current $\Jcal$,
that is when \hbox{$\Jcal=i_X\Lcal+i_W\Pcal$} with
\hbox{$X:\JO\E\to\TO\M$}\,, \hbox{$W:\JO\E\to\VO\E$}.
It's easy to check that in that case one such $Y$ is obtained by setting
$$Y=X\pint\mathrm{d\!l}+W=X^a\,\de\xx_a+(X^a\,\yy^i_a+W^i)\,\de\yy_i~,$$
where \hbox{$X=X^a\,\de\xx_a$}\,, \hbox{$W=W^i\,\de\yy_i$}\,.
We'll actually use this procedure
in~\sref{ss:The off-shell symmetry associated with any vector field}.

\begin{remark}
In the physics literature the above topics are usually expressed
in terms of ``covariant divergence'' rather than of exterior differential.
The relation between these two formalisms, which we'll further discuss
in~\sref{ss:Remarks about the replacement principle},
is also called the ``replacement principle''~\cite{Janyska07}.
\end{remark}

\subsection{Canonical energy-tensor in non-trivial bundles}
\label{ss:Canonical energy-tensor in non-trivial bundles}

We are still in the context of a 1-st order Lagrangian field theory
on a fibered manifold \hbox{$\E\onto\M$},
but now we also assume a (provisionally fixed)
connection \hbox{$\k:\E\to\JO\E$}.
We can then define the \emph{canonical energy-tensor} as the morphism
\hbox{$\Ucal:\JO\E\to\weu{m-1}\TS\M\tn\TS\M$} over $\M$
which has the coordinate expression
$$\Ucal=\bigl(\ell\,\d\Ii ab-(\yy^i_b-\k^i_b)\,\de^a_i\ell\bigr)\,
\dx_a\tn\dx^b~.$$
The importance of this object lies in the possibility to consider
certain symmetries which are generated by vector fields on the base manifold.
Actually for any section \hbox{$\phi:\M\to\E$}
and for any vector field \hbox{$X:\M\to\TO\M$} we have
$$(\Ucal\comp\jO\phi)\pint X=\jO\phi^*(i_Y\Ccal)~,$$
where \hbox{$Y\equiv X\pint\k:\E\to\TO\E$}
is the so-called \emph{horizontal prolongation} of $X$.
Thus, assuming that $Y$ turns out to be a symmetry of $\Ccal$,
the conserved current evaluated through a critical section $\phi$ is
\hbox{$(\Ucal\comp\jO\phi)\pint X:\M\to\weu3\TS\M$}.

We briefly review the geometric construction of $\Ucal$~\cite{CM85}.
The covariant derivative operator associated with $\k$ can be regarded
as a morphism \hbox{$\nabla:\JO\E\to\TS\M\ten{\E}\VO\E$}.
By performing suitable contractions of the tensor product $\DO\Lcal\tn\nabla$
we obtain the morphism
$$\bang{\DO\Lcal\tn\nabla}=(\yy^i_a-\k^i_a)\,\de^b_i\ell\,\dx^a\tn\dx_b:
\JO\E\to\weu{m-1}\TS\M\ten{\M}\TS\M~.$$
Moreover we note that there is a natural inclusion
\hbox{$\imath:\weu{m}\TS\M\into\weu{m-1}\TS\M\ten{\M}\TS\M$},
characterized by \hbox{$i_X\a=(\imath\a)\pint X$}\,,
so that eventually we set \hbox{$\Ucal:=\imath\Lcal-\bang{\DO\Lcal\tn\nabla}$}.

\smallbreak
In order to compare the above $\Ucal$
with the stress-energy tensor $\Tcal$ of standard General Relativity we note that the latter
is valued into \hbox{$\TO\M\ten{\M}\TO\M\ten{\M}\weu4\TS\M$};
indeed, when the matter Lagrangian density $\Lcal$ does not depend
on the derivatives of the metric $g$,
then $\Tcal$ is just the fiber derivative of $\Lcal$ with respect to $g$.
By a contraction and index moving via $g$ we obtain a morphism
$$\Tcal=\Tcal\Ii ab\,\dx_a\tn\dx^b:\JO\E\to\weu3\TS\M\ten{\M}\TS\M~.$$
So $\Ucal$ and $\Tcal$ are valued in the same bundles,
and it's easy to see that their ``physical dimensions'' match too.
While they not necessarily coincide for an arbitrary Lagrangian density,
in physically relevant cases they turn out to be either equal
or closely related.\footnote{
See e.g.\ Gotay-Marsden~\cite{GotayMarsden92} for a discussion
about relations between these two types of objects.} 

\subsection{Remarks about the ``replacement principle''}
\label{ss:Remarks about the replacement principle}

If the base manifold of a field theory is Lorentz spacetime,
with positively oriented unit volume form $\eta$\,,
then a current \hbox{$\Jcal=\Jcal^a\,\dx_a$} can be treated as the vector field
\hbox{$J:\M\to\TO\M$} characterized by \hbox{$\Jcal=i_{{\!}_\sJ}\eta$}\,,
that is \hbox{$J=J^a\,\de\xx_a$} with \hbox{$J^a\,\rdg=\Jcal^a$}\,.
Accordingly, the exterior differential $\dO\Jcal$ can be replaced by
the covariant ``divergence'' $\na_aJ^a$,
as it's not difficult to check that if the spacetime connection is torsion-free
then actually \hbox{$\dO\Jcal=\na_aJ^a\,\rdg\,\dO^4\xx$}\,,
so that the two formalisms could be easily merged.

We can extend the above procedure as follows.
We assume that \hbox{$\E\onto\M$} is a vector bundle
(hence \hbox{$\VO\E\cong\E\cart{\M}\E$}).
We consider a linear connection \hbox{$\k:\E\to\JO\E$}
besides the spacetime connection $\G$, and a section
$$\xi=\xi^{ai}\,\dx_a\tn\de\yy_i:\M\to\weu3\TS\M\ten{\E}\VO\E~.$$
Then we have the Fr\"olicher-Nijenhuis bracket
$$\fnb{\k,\xi}=(\de_a\xi^{ai}-\k\iIi aij\,\xi^{aj})\,\dO^4\xx\tn\de\yy_i\;:\;
\M\to\weu4\TS\M\ten{\E}\VO\E~.$$

Moreover we have the covariant derivative of $\xi$ with respect to $(\G,\k)$,
with the coordinate expression
$$\nabla\xi=(\de_c\xi^{ai}-\G\iIi cab\,\xi^{bi}+\G\iIi cbb\,\xi^{ai}
-\k\iIi cij\,\xi^{aj})\,\dx^c\tn\dx_a\tn\de\yy_i~.$$
As this is valued in $\TS\M\ten{\M}\weu3\TS\M\ten{\E}\VO\E$
we can antisymmetrize the horizontal factors,
thus obtaining the \emph{covariant divergence}
\begin{align*}
\codiv\xi&=(\de_a\xi^{ai}-\k\iIi aij\,\xi^{aj}
+\t_a\,\xi^{ai})\,\dO^4\xx\tn\de\yy_i=
\\[6pt]
&=\fnb{\k,\xi}+\t\we\xi~,
\end{align*}
where \hbox{$\t_a\equiv \G\iIi acc-\G\iIi cca$} is the \emph{torsion $1$-form}.

Via natural algebraic operations involving the volume form $\eta$
we also introduce
$$\breve\xi=\breve\xi{}^{ai}\,\de\xx_a\tn\de\yy_i:\M\to\TO\M\ten{\M}\E~,\qquad
\breve\xi{}^{ai}\equiv\tfrac1\rrdg\,\xi^{ai}~,$$
and by a straightforward computation we find
$$\codiv\xi=\codiv\breve\xi\tn\eta~,$$
where
$$\codiv\breve\xi\equiv\nabla_a\breve\xi{}^{ai}\,\de\yy_i=
\tfrac1\rrdg\,(\de_a\xi{}^{ai}-\k\iIi aij\,\xi^{aj}+\t_a\,\xi^{ai})\,\de\yy_i~.$$

For handling energy-tensors we just set \hbox{$\E\equiv\TS\M$}, and get
$$\codiv\,\Ucal=\codiv\,\breve\Ucal\tn\eta=
\na_a\breve\Ucal\Ii ab\,\rdg\,\dx^b\tn\dO^4\xx~,\qquad
\breve\Ucal\Ii ab\equiv\tfrac1\rrdg\,\Ucal\Ii ab~.$$

\subsection{Basic examples}
\label{ss:Basic examples}

In concrete examples,
particularly when the theory under consideration has several sectors,
dropping the rigorous distinction between fiber coordinates and field components
can be notationally convenient,
and usually won't generate confusion if some care is used.
Accordingly we'll also write $\Tcal$ and $\Ucal$
for $\Tcal\comp\jO\phi$ and $\Ucal\comp\jO\phi$\,.
The expressions of the energy tensors and their covariant divergences
found in the two following examples are essentially standard results,
though we stress that, differently from usual presentations,
a field and its conjugate are seen here
as independent sections of mutually dual bundles.
The most important point about these examples is their association with the
energy-tensor of a gauge field (\sref{ss:Energy-tensor of a gauge field}).

\subsubsection{Charged spin-zero field (non-abelian case)}
\label{sss:Charged spin-zero field}

In this example \hbox{$\F\onto\M$} is a vector bundle
endowed with a linear connection $\k$\,,
and $(\M,g)$ is Einstein spacetime.
Provisionally, both $\k$ and $g$ are considered as fixed background structures.
The ``configuration bundle'' is \hbox{$\E\equiv\F\dir{\M}\F^*$},
so that a field is actually a couple of sections,
\hbox{$\phi:\M\to\F$} and \hbox{$\bar\phi:\M\to\F^*$}.
In standard presentations the fibers are complex
and $\phi$ and $\bar\phi$ are regarded as mutually adjoint
via some Hermitian structure preserved by $\k$\,,
but that specification is not needed here.

The Lagrangian density \hbox{$\Lcal=\ell\,\dO^4\xx$} is given by
$$\ell=\oh\,(g^{ab}\,\na_a\bar\phi_i\,\na_b\phi^i-m^2\,\bar\phi_i\,\phi^i)\,\rdg~,$$
where \hbox{$\na_a\phi^i=\de_a\phi^i-\k\iIi aij\,\phi^j$}\,,
\hbox{$\na_a\bar\phi_i=\de_a\bar\phi_i+\k\iIi aji\,\bar\phi_j$}\,,
and $m$ is a constant mass.

We have
$$\Pcal^a_i\equiv\de^a_i\ell=\oh\,g^{ac}\,\na_c\bar\phi_i~,\qquad
\Pcal^{ai}\equiv\de^{ai}\ell=\oh\,g^{ac}\,\na_c\phi^i~,$$
whence we obtain
\begin{align*}
\Ucal\Ii ab&=\ell\,\d\Ii ab-\Pcal^a_i\,\na_{b}\phi^i-\na_{b}\bar\phi_i\,\Pcal^{ai}=
\\[6pt]
&=\oh\,\bigl(g^{cd}\,\na_c\bar\phi_i\,\na_d\phi^i\,\d\Ii ab
-g^{ac}\,(\na_c\bar\phi_i\,\na_b\phi^i+\na_b\bar\phi_i\,\na_c\phi^i)
-m^2\,\bar\phi_i\,\phi^i\,\d\Ii ab\bigr)\,\rdg~,
\end{align*}

Next, using \hbox{$\de\rdg/\de g^{ab}=-\oh\,g_{ab}\,\rdg$}\,,
we obtain
\begin{align*}
\Tcal_{ab}&=\oq\,\na_{\{a}\bar\phi_i\,\na_{b\}}\phi^i\,\rdg-\oh\,g_{ab}\,\ell=
\\[6pt]
&=\oq\,(\na_a\bar\phi_i\,\na_b\phi^i+\na_b\bar\phi_i\,\na_a\phi^i)\rdg
-\oq\,g_{ab}\,(g^{cd}\,\na_c\bar\phi_i\,\na_d\phi^i-m^2\,\bar\phi_i\,\phi^i)\,\rdg~,
\end{align*}
where braces delimiting indices denote symmetrization
(without dividing by the appropriate factorial).
By a simple further calculation we then obtain
$$g_{ac}\,\Ucal\Ii cb=-2\,\Tcal_{ab}~.$$

We evaluate $\codiv\,\Ucal$ on-shell,
that is by taking the field equations into account.
These can be expressed in terms of the Fr\"olicher-Nijenhuis bracket in the form
$$\fnb{\k,{*}\nabla\phi}+m^2\,\phi\,\eta=0~,\qquad
\fnb{\k,{*}\nabla\bar\phi}+m^2\,\bar\phi\,\eta=0~,$$
where ${*}$ is the standard Hodge isomorphism, or in coordinates as
\begin{align*}
&\de_a(g^{ab}\,\rdg\,\na_b\phi^i)-g^{ab}\,\rdg\,\k\iIi aij\,\na_b\phi^j
+m^2\rdg\,\phi^i=0~,
\\[6pt]
&\de_a(g^{ab}\,\rdg\,\na_b\bar\phi_i)+g^{ab}\,\rdg\,\k\iIi aji\,\na_b\bar\phi_j
+m^2\rdg\,\bar\phi_i=0~.
\end{align*}
A computation then yields the on-shell expression
$$\na_a\breve\Ucal\Ii ab=
\oh\,g^{ac}\,\r\iIi{ab}ij\,(\bar\phi_i\,\na_c\phi^j-\na_c\bar\phi_i\,\phi^j)~.$$

\subsubsection{Dirac field}\label{sss:Dirac field}

Let \hbox{$\W\onto\M$} be the bundle of Dirac spinors over Einstein spacetime.
In this context we indicate by \hbox{$\g:\TO\M\to\End\W$} the Dirac map,
while the background \emph{spinor connection} splits as
$$\Cs\iIi a\a\b=\iO\,A_a\,\d\Ii\a\b+\oq\,\G\iI a{\l\m}\,(\g_\l\,\g_\m)\Ii\a\b~,$$
where $A$ is real and represents the e.m.\ field
and $\G$ is the spacetime connection,
whose components are expressed here
in an orthonormal frame $\bigl(\theta_\l\bigr)$ of \hbox{$\TO\M\onto\M$}
(a ``tetrad'').
As in the previous example we consider two independent fields
\hbox{$\psi:\M\to\W$} and \hbox{$\bar\psi:\M\to\W^*$},
which are usually regarded as mutually related by the \emph{Dirac adjunction},
determined by a Hermitian metric with signature $(++--)$.

We set
\begin{align*}
\ell&=\bigl(\ih\,(\bar\psi\,\nasl\!\psi-\nasl\!\bar\psi\,\psi)
-m\,\bar\psi\,\psi\bigr)\,\rdg\equiv
\\[6pt]
&\equiv\bigl(\ih\,g^{ab}\,(\bar\psi_\a\,\g\iIi a\a\b\,\na_b\psi^\b
-\na_a\bar\psi_\a\,\g\iIi b\a\b\,\psi^\b)
-m\,\bar\psi_\a\,\psi^\a\bigr)\,\rdg~,
\end{align*}
where \hbox{$\na_a\psi^\a=\de_a\bar\psi_\a-\Cs\iIi a\a\b\,\psi^\b$}\,,
\hbox{$\na_a\bar\psi_\a=\de_a\bar\phi_\a+\Cs\iIi a\b\a\,\bar\psi_\b$}\,.
Then
\begin{align*}
&\Tcal_{ab}=\frac{\de\ell}{\de g^{ab}}=
\iq\,(\bar\psi_\a\,\g\iIi{\{a}\a\b\,\nabla_{\!b\}}^{\phantom{A}}\psi^\b
-\nabla_{\!\!\{a}^{\phantom{A}}\bar\psi_\a\,\g\iIi{b\}}\a\b\,\psi^\b)\,\rdg
-\oh\,\ell\,g_{ab}\equiv
\\[6pt]&\phantom{ \Tcal_{ab} }
\equiv \iq\,\bigl(
\bar\psi\,\g_{\{a}^{\phantom{A}}\nabla_{\!b\}}^{\phantom{A}}\psi
-\nabla_{\!\{a}^{\phantom{A}}\bar\psi\,\g_{b\}}^{\phantom{A}}\psi\bigr)\,\rdg
-\oh\,\ell\,g_{ab}~.
\end{align*}

Moreover we have
$$\Pcal^a_\a\equiv\de^a_\a\ell=\ih\,(\bar\psi\,\g^a)_\a\,\rdg~,\qquad
\Pcal^{a\a}\equiv\de^{a\a}\ell=-\ih\,(\g^a\,\psi)^\a\,\rdg~,$$
whence
\begin{align*}
\Ucal\Ii ab&=
-\ih\,\bigl(\bar\psi\,\g^a\,\na_b\psi-\na_b\bar\psi\,\g^a\,\psi\bigr)
+\ell\,\d\Ii ab~,
\\[10pt]
\Ucal_{ab}&=
-\ih\,\bigl(\bar\psi\,\g_a\na_b\psi-\na_b\bar\psi\,\g_a\psi\bigr)
+\ell\,g_{ab} \qR
\\[10pt] \Rq
\oh\,\Ucal_{\{ab\}}&=
-\ih\,\bigl(
\bar\psi\,\g_{\{a}^{\phantom{A}}\nabla_{\!b\}}^{\phantom{A}}\psi
-\nabla_{\!\{a}^{\phantom{A}}\bar\psi\,\g_{b\}}^{\phantom{A}}\psi\bigr)\,\rdg
+\ell\,g_{ab}=
\\[6pt]
&=-2\,\Tcal_{ab}~.
\end{align*}

Next we evaluate the divergence of $\Tcal$ on-shell,
that is by taking the Dirac equations
$$\g^a\na_a\psi=-\iO\,m\,\psi~,\qquad \na_a\bar\psi\,\g^a=\iO\,m\,\bar\psi~,$$
into account (for simplicity we are considering the case
when the torsion of the spacetime connection vanishes).
A not-so-short computation then yields
$$\na_a\breve\Tcal\Ii ab=\ih\,\r_{ab}\,\bar\psi\,\g^a\,\psi=
\oh\,F_{ab}\,\bar\psi\,\g^a\,\psi~.$$

\section{Energy-tensor of a connection}
\label{s:Energy-tensor of a connection}

Henceforth, as in the previous examples,
we'll consistently use the notational simplification of dropping the distinction
between fiber coordinates and field components. In the literature,
the adjective ``formal'' is sometimes attached to maps defined
on jet bundles in a setting of this type;
so for example one writes ``formal curvature''~\cite{Krupka02} and the like.

\subsection{Energy-tensor of a gauge field}
\label{ss:Energy-tensor of a gauge field}

We now use results of previous sections
in order to propose a definition of energy-tensor of a gauge field.
It's worthwhile stressing how this object naturally associates itself
with the energy-tensors of matter fields~(\sref{ss:Basic examples}).

Let $\M$ be a spacetime with fixed background metric.
In a theory of a gauge field \hbox{$\k:\M\to\C\!_\Gg$}
(\sref{ss:Systems of gauge fields}) one uses the
natural Lagrangian \hbox{$\Lcal\spec{gauge}=\ell\spec{gauge}\,\dO^4\xx$} with
$$\ell\spec{gauge}=-\oq\,g^{ac}\,g^{bd}\,\ost\r_{ab\sI}\,\r\iI{cd}\sI\,\rdg~.$$
Since we also have the background Riemannian connection
associated with the spacetime metric,
a gauge field yields an overconnection $\k^\up$
which is reducible to a connection of \hbox{$\C\!_\Gg\onto\M$}:
in its expression given in~\sref{ss:Systems of gauge fields}
we just replace the generic $\G$ with the Levi-Civita connection.
Thus we are naturally led use $\k^\up$ in order to extend
the construction of the canonical energy-tensor
offered in~\sref{ss:Canonical energy-tensor in non-trivial bundles}:
we insert the covariant derivative of $\k$ with respect to $\k^\up$
in the place that there is occupied by the covariant derivative
of the field with respect to the background connection.
While this is not exactly the same procedure,
it will be justified by the properties obeyed by the considered object.

In relation to the Utiyama theorem~\cite{Janyska07} we note
that viewing $\r$ as the covariant derivative of $\k$ puts
gauge and matter fields on a more similar footing:
the derivatives of both fields enter the total Lagrangian
only through the covariant derivatives.

The momentum components for $\k$ and its covariant derivative are
$$\Pcal^a{}^c_\sI\equiv\dde{\ell\spec{gauge}}{\k^\sI_{c,a}}
=\ost\r\Ii{ac}\sI~,\qquad
\na_b\k_c^\sI=-\r\iI{bc}\sI~,$$
whence by applying the above sketched procedure we get
the canonical energy-tensor $\Ucal\spec{gauge}$ with components
$$\Ucal\Ii ab=\ell\,\d\Ii ab-\Pcal^a{}^c_\sI\,\na_b\k_c^\sI=
\bigl(\ost\r\Ii{ac}\sI\,\r\iI{bc}\sI
-\oq\,\ost\r\Ii{cd}\sI\,\r\iI{cd}\sI\,\d\Ii ab\bigr)\rdg~.$$
When \hbox{$\k\equiv\iO\,A$} where $A$ the electromagnetic 4-potential,
$\Ucal\spec{gauge}$ is seen to coincide with the Maxwell stress-energy tensor.

We also note that, analogously to the examples in~\sref{ss:Basic examples},
we find
$$-\oh\,\Ucal_{ab}=\Tcal_{ab}\equiv \de\ell/\de g^{ab}~.$$

Let now \hbox{$X:\M\to\TO\M$} and denote by
$$Y\equiv X\pint\k^\up=
X^c\,(\de\xx_c+(\k^\up_c)^i_{bj}\,\de\yy_i^{bj}):\C\!_\Gg\to\TO\C\!_\Gg$$
its horizontal lift through $\k^\up$.
We use coordinates $\bigl(\xx^a,\yy^\sI_a\,,\yy^\sI_{a,b}\bigr)$ on $\JO\C\!_\Gg$\,,
and write
$$\Pcal\!\!\spec{gauge}=\Pcal^a{}^c_\sI\,\vartheta^\sI_c\we\dx_a\equiv
\Pcal^a{}^c_\sI\,(\dy^\sI_c-\yy^\sI_{c,b}\,\dx^b)\we\dx_a~.$$
Then by a coordinate computation it's not difficult to prove:
\begin{theorem}
For each section \hbox{$\k:\M\to\C\!_\Gg$} we have
$$(\Ucal\spec{gauge}\comp\jO\k)\pint X=\jO\k^*(i_Y\Ccal\spec{gauge})~,\qquad
\Ccal\spec{gauge}\equiv\Lcal\spec{gauge}+\Pcal\!\!\spec{gauge}~.$$
\end{theorem}

The above theorem constitutes a first justification for our calling
$\Ucal\spec{gauge}$ the canonical energy-tensor of the gauge field.
We can offer a further justification,
related to the vanishing of the divergence of energy tensors.
Indeed it can be shown by a general naturality argument~\cite{LandauLifchitz68,HE} that
the energy-tensor for a field theory on a gravitational background
must be divergence-free.
In order to check that this requirement is fulfilled
we need to consider a theory of interacting matter and gauge fields
$$(\phi,\k):\M\to\E\cart{\M}\C\!_\Gg~.$$
As a typical basic example, we first consider the charged boson field
of~\sref{sss:Charged spin-zero field}
together with the appropriate gauge field.
The Lagrangian density for this theory is assumed to be
just the sum \hbox{$\Lcal=\Lcal_\phi+\Lcal\spec{gauge}$}\,.
Note that $\Lcal_\phi$ depends on $\k$ but not on its derivatives,
while $\Lcal\spec{gauge}$ is independent of $\phi$\,.
Then also the total momentum is the sum
\hbox{$\Pcal=\Pcal\!\!_\phi+\Pcal\!\!\spec{gauge}$}\,,
and the total energy-tensor is the sum
\hbox{$\Ucal=\Ucal_\phi+\Ucal\spec{gauge}$}\,.
The field equations for $\k$ can be expressed
in terms of the Fr\"olicher-Nijenhuis bracket in the form
$$\fnb{\ost\k,{*}\ost\r}\IiI aij+
\oh\,g^{ab}\,(\bar\phi_i\,\na_b\phi^j-\na_b\bar\phi_i\,\phi^j)=0~,$$
whence we get the on-shell divergence
$$\codiv(\breve\Ucal\spec{gauge})_b=
\oh\,g^{ac}\,(\bar\phi_i\,\na_c\phi^j-\na_c\bar\phi_i\,\phi^j)\,\r\iIi{ba}ij~.$$
Thus eventually, taking the replacement principle into account, we find
$$\codiv(\Tcal_\phi+\Tcal\!\!\spec{gauge})=0~.$$

Analogous computations yield the same result in the case of a theory
of interacting fermion and gauge fields.
Restricting our attention to the abelian case for simplicity,
the field equations for $\k$ and the on-shell divergence
of $\Ucal\spec{gauge}$ are now
$$\fnb{\ost\k,{*}\ost\r}^a-\iO\,\bar\psi\,\g^a\,\psi=0~,\qquad
\na_a\breve\Ucal\Ii ab[\k]=\iO\,\bar\psi\,\g^a\,\psi\,\r_{ab}~.$$
Then \hbox{$\codiv(\Tcal_\psi+\Tcal\!\!\spec{gauge})=0$}\,,
where $\Tcal_\psi$ is related to $\Ucal_\psi$ by symmetrization
(\sref{sss:Dirac field}).

\subsection{Energy-tensor of the gravitational field}
\label{ss:Energy-tensor of the gravitational field}

A convenient Lagrangian formulation of gravity,
discussed in the literature, is obtained by letting
the spacetime metric $g$ and the symmetric spacetime connection $\G$
be independent variables,
and assuming the Lagrangian density to be
\hbox{$\Lcal\spec{grav}=R\,\rdg\,\dO^4\xx$} where
\hbox{$R\equiv g^{ac}\,R\iIi{ab}bc$} denotes the scalar curvature.
The Euler-Lagrange equations turn out to be the Einstein equation for the $g$ sector,
and the metricity condition \hbox{$\nabla[\G]g=0$} for the $\G$ sector.

Let \hbox{$\C_{\!\sst\G}\onto\M$} be the bundle
of symmetric linear connections of \hbox{$\TO\M\onto\M$}.
Then any \hbox{$\G:\M\to\C_{\!\sst\G}$} determines an overconnection
\hbox{$\G^\up:\C_{\!\sst\G}\to\JO\C_{\!\sst\G}$}\,;
we don't have to worry about an auxiliary connection on the base manifold
since we avail of $\G$ itself.
We can now proceed analogously to~\sref{ss:Energy-tensor of a gauge field}.
The ``covariant derivative'' of $\G$ with respect to $\G^\up$
is \hbox{$\na_a\G\iIi bcd=-R\iIi{ab}cd$}\,,
and since the Lagrangian is independent of the derivatives of the metric
the momentum map $\Pcal\!\!\spec{grav}$ has the simple coordinate expression
\begin{align*}
&\Pcal\!\!\spec{grav}=\Pcal\IiI{ab}cd\,(\dy\iIi bcd\we\dx_a-\yy\iIi bc{d,a}\,\dO^m\xx)~,
\\[8pt]
&\Pcal\IiI{ab}cd=(-g^{ad}\,\d^b_c+g^{bd}\,\d^a_c)\,\rdg~.
\end{align*}
By the way, we remark that the above expression is frequent in physics.
In particular, it is related to coefficients $Q\Ii{ab}{cd}$
used e.g.\ by Padmanabhan~\cite{Padmanabhan1312.3253} by
\hbox{$\Pcal\Ii{ab}{cd}=2\,\rdg\,Q\Ii{ab}{cd}$}\,.

We also remark that Krupka has studied energy-tensors
in terms of variational forms~\cite{Krupka02}.
In that approach the gravitational field is represented by the metric alone
(a choice which makes certain details a little more involved).

The energy-tensor turns out to be essentially the Einstein tensor $G$,
since we get
\begin{align*}
(\Ucal\spec{grav}\!)\Ii ab&=R\,\rdg\,\d\Ii ab-
\na_b\G\iIi ecd\,\Pcal\IiI{ae}cd=
\bigl(R\,\d\Ii ab+R\iIi{be}cd\,(g^{ad}\,\d^e_c-g^{ed}\,\d^a_c)\bigr)\,\rdg=
\\[6pt]
&=-2\,(R\Ii ab-\oh\,R\,\d\Ii ab)\,\rdg\equiv-2\,G\Ii ab\,\rdg~.
\end{align*}

The horizontal $\G^\up$-lift of \hbox{$X:\M\to\TO\M$} is the vector field
\hbox{$X\pint\G^\up:\C_{\!\sst\G}\to\TO\C_{\!\sst\G}$}.
Remembering that \hbox{$\na_aG\Ii ab=0$}\,,
and letting \hbox{$\Ccal\spec{grav}=\Lcal\spec{grav}+\Pcal\!\!\spec{grav}$}
be the Poincar\'e-Cartan form for this gravitational setting,
a short computation also yields
$$\jO\G^*\dO(i_{X\pint\G^\up}\Ccal\spec{grav})=\bang{\Ucal\spec{grav}\,,\nabla X}=
-2\,G\Ii ab\,\na_bX^a\,\rdg\,\dO^4\xx~.$$

Let's now consider a theory of interacting matter, gauge and gravitational fields,
and the total energy-tensor
\hbox{$\Tcal=\Tcal\!\!\spec{matter}+\Tcal\!\!\spec{gauge}+G=-\oh\,\Ucal$}\,.
Then the condition that $\Tcal$ vanishes is equivalent to the Einstein equations
$$G=-(\Tcal\!\!\spec{matter}+\Tcal\!\!\spec{gauge})~.$$

The total bundle \hbox{$\Z\onto\M$} for this theory has a  matter sector,
a gauge sector, and a gravitational sector which in turn has a $g$ sector
and a $\G$ sector.
Gathering the various constructions we readily realize that the couple $(\k,\G)$\,,
constituted by a gauge field and a spacetime connection,
determines a prolongation of any basic vector field \hbox{$X:\M\to\TO\M$}
to a vector field \hbox{$Y[X]:\Z\to\TO\Z$}.
Now, knowing that \hbox{$\Tcal\!\!\spec{matter}+\Tcal\!\!\spec{gauge}$}
is identically divergence-free on-shell,
we can reformulate a known result~\cite{Padmanabhan1312.3253}
in terms of symmetries of the \emph{total} Poincar\'e-Cartan form
\hbox{$\Ccal=\Ccal\spec{grav}+\Ccal\spec{matter}+\Ccal\spec{gauge}$}\,:
\begin{theorem}
The Einstein field equations follow from the requirement
that $i_{Y[X]}\Ccal$ be a conserved current
for every basic vector field $X$.\end{theorem}

\subsection{The off-shell symmetry associated with any vector field}
\label{ss:The off-shell symmetry associated with any vector field}

If $\M$ is an $m$-dimensional manifold
then any exterior form \hbox{$\varphi:\M\to\weu{m-2}\TS\M$}
trivially yields the conserved current $\dO\varphi$
for any field theory in which $\M$ is the ``source''.
In this section $\M$ is the spacetime manifold and the Riemannian connection
is assumed to be torsion-free.
Then for any $2$-form $\varphi$ we also consider
the current \hbox{$\oh\dO{*}\varphi$}
(${*}$ is the Hodge isomorphism).
This can be expressed by a version of the replacement principle
(\sref{ss:Poincare-Cartan form and currents})
in terms of the covariant divergence, since we have
$$\oh\dO{*}\varphi=\na_a\varphi^{ab}\,\rdg\,\dx_b~.$$

Hence we can associate a current
to any vector field \hbox{$X:\M\to\TO\M$} by setting
$$\Jcal:=\oh\dO{*}\dO[g^\flat(X)]~,\qquad
g^\flat(X)\equiv g_{ac}\,X^c\,\dx^a:\M\to\TS\M~.$$
Since the torsion is assumed to vanish we also have
$$\oh{*}\dO[g^\flat(X)]_{ab}=\oh\,\na_{[a}X_{b]}\,\rdg~,$$
which is known in the literature as the
``Komar superpotential''~\cite{Komar59,Padmanabhan1312.3253}.
Now we can rewrite the current as
$$\Jcal=\na_a\nabla^{[a}X^{b]}\,\rdg\,\dx_b\equiv J^b\,\rdg\,\dx_b~,$$
and its closeness can be expressed as \hbox{$\na_bJ^b=0$}\,.
We remark that this is an \emph{off-shell} symmetry,
namely it holds independently of the field equation
possibly obeyed by the gravitational field.

While this current is associated with a vector field on $\M$,
it is not obtained as $\Ucal\pint X$,
where \hbox{$\Ucal\equiv\Ucal\spec{grav}=-2\,G$}
(\sref{ss:Energy-tensor of the gravitational field}); namely
if \hbox{$\G:\M\to\C_{\!\sst\G}$} is an arbitrary section,
then the contraction of the horizontal lift
\hbox{$X\pint\G^\up:\C_{\!\sst\G}\to\TO\C_{\!\sst\G}$}
with the Poincar\'e-Cartan form $\Ccal\spec{grav}$ does not yield
the conserved current $\Jcal$.
Hence it is natural to look for a different lift $Y$ of $X$
such that \hbox{$\Jcal\comp\jO\G=\jO\G^*(i_Y\Ccal\spec{grav})$}\,.
We'll use a procedure sketched in~\sref{ss:Poincare-Cartan form and currents}.

We first recall that the
\emph{Lie derivative of the connection $\G$ with respect to $X$}
is the tensor field \hbox{$\LO_X\G:\M\to\TS\M\tn\TO\M\tn\TS\M$}
characterized by~\cite{Yano55}
$$\LO_X\G\pint Z=\nabla\LO_XZ-\LO_X\nabla Z$$
for any vector field \hbox{$Z:\M\to\TO\M$}.
We obtain the coordinate expression
\begin{align*}
\LO_X\G\iIi abc&=
-\de_{ac}X^b+\de_aX^d\,\G\iIi dbc+\G\iIi abd\,\de_cX^d
-\G\iIi adc\,\de_d X^b+X^d\,\de_d\G\iIi abc=
\\&=-\na_a\na_cX^b-X^d\,R\iIi{da}bc~,
\end{align*}
whence
$$J^b=\na_a\nabla^{[a}X^{b]}=
-g^{ac}\,\LO_X\G\iIi abc+g^{ab}\,\LO_X\G\iIi acc+2\,R\Ii ba\,X^a~.$$

The Poincar\'e-Cartan form for this case is
\hbox{$\Ccal\spec{grav}\equiv\Lcal\spec{grav}+\Pcal\!\!\spec{grav}$}\,,
already written in~\sref{ss:Energy-tensor of the gravitational field}.
Then one immediately checks that the current's components can be rewritten as
$$J^a=-\Pcal\IiI{ab}cd\,\LO_X\G\iIi bcd+2\,R\Ii{a}{b}\,X^b~.$$

If now \hbox{$Y=Y^a\,\de\xx_a+Y\iIi abc\,\de\yy\IiI abc:
\JO\C_{\!\sst\G}\to\TO\C_{\!\sst\G}$}
is a morphism over $\C_{\!\sst\G}$\,, then we obtain
$$\jO\G^*(i_Y\Ccal\spec{grav})= \bigl( R\,Y^a
+\Pcal\IiI{ab}cd\,(Y\iIi bac-Y^d\,\de_d\G\iIi bac)\bigr)\,\rdg\,\dx_a~.$$
The condition
\hbox{$\jO\G^*(i_Y\Ccal\spec{grav})=\Jcal\comp\jO\G$} can be expressed as
$$R\,Y^a+\Pcal\IiI{ab}cd\,(Y\iIi bcd-\G\iIi bc{d,e}\,Y^e)=
-\Pcal\IiI{ab}cd\,\LO_X\G\iIi bcd+2\,R\Ii{a}{b}\,X^b~.$$
Since we are looking for \emph{one} solution $Y$,
the first obvious assumption is that $Y$ projects over $X$, that is \hbox{$Y^a=X^a$}.
Furthermore, by a straightforward computation we also get
$$\Pcal\IiI{ab}cd\,(R\iIi{bd}ce\,X^e-R_{bd}\,X^c)=2\,R\Ii{a}{b}\,X^b-R\,X^a~,$$
so that eventually we are led to write the equation
$$\Pcal\IiI{ab}cd\,(Y\iIi bcd-X^e\,\de_e\G\iIi bc{d}+\LO_X\G\iIi bcd
-R\iIi{bd}ce\,X^e+R_{bd}\,X^c)=0$$
in which the unknowns are the components $Y\iIi bcd$\,.
One solution is
$$Y\iIi bcd=X^e\,\de_e\G\iIi bc{d}-\LO_X\G\iIi bcd
+R\iIi{bd}ce\,X^e-R_{bd}\,X^c~.$$

Summarizing, and writing our result in a somewhat more precise form:
\begin{theorem}
For any vector field \hbox{$X:\M\to\TO\M$}, the morphism
\begin{align*}
&Y=X^a\,\de\xx_a+Y\iIi bcd\,\de\yy\IiI bcd:\JO\C_{\!\sst\G}\to\TO\C_{\!\sst\G}~,
\\[6pt]
&Y\iIi bcd\comp\jO\G=X^e\,\de_e\G\iIi bc{d}-\LO_X\G\iIi bcd
+R\iIi{bd}ce[\G]\,X^e-R_{bd}[\G]\,X^c~,
\end{align*}
yields the standard current $\Jcal[X]$ associated with $X$ as $i_Y\Ccal$\,.
\end{theorem}
\remark~In coordinate-free form we may express $Y$ as
$$Y=X\pint\mathrm{d\!l}
-\LO_X\G+\mathrm{Riemann}\pint X-\mathrm{Ricci}\tn X~,$$
where the last three terms are valued in
\hbox{$\VO\C_{\!\sst\G}\cong\C_{\!\sst\G}\cart{\M}\TS\M\tn\TO\M\tn\TS\M$}.


\end{document}